\documentclass[a4paper]{jpconf}

\usepackage{graphicx}
\usepackage{iopams}
\usepackage{hyperref}
\usepackage{bm}

\newcommand{\meV}{{\rm meV}}
\newcommand{\eV}{{\rm eV}}

\newcommand{\MeV}{{\rm MeV}}
\newcommand{\GeV}{{\rm GeV}}
\newcommand{\TeV}{{\rm TeV}}

\newcommand{\Mpl}{M_{\rm Pl}}

\newcommand{\nubar}{\bar{\nu}}

\begin{document}
\title{Quantum Gravity phenomenology:\\ achievements and challenges}

\author{S Liberati$^{1,2}$ and L Maccione$^{3}$}

\address{$^{1}$ SISSA, via Bonomea, 265, 34136 Trieste, Italy}
\address{$^{2}$ INFN, Sezione di Trieste, Via Valerio 2, 34127 Trieste, Italy}
\address{$^{3}$ DESY, Theory Group, Notkestra{\ss}e 85, 22607 Hamburg, Germany}

\ead{liberati@sissa.it}
\ead{luca.maccione@desy.de}

\begin{abstract}
Motivated by scenarios of quantum gravity, Planck-suppressed deviations from Lorentz invariance are expected at observable energies. Ultra-High-Energy Cosmic Rays, the most energetic particles ever observed in nature, yielded in the last two years strong constraints on deviations suppressed by $O(E^{2}/\Mpl^{2})$ and also, for the first time, on space-time foam, stringy inspired models of quantum gravity. We review the most important achievements and discuss future outlooks.
\end{abstract}

\section{Introduction}

Quantum Gravity (QG) has posed a challenge to many theoretical physicists of the last generation and is at present far from understood. Although we do not yet have a single experiment or observation forcing us to introduce such a theory\footnote{However, part of the gravitation theory community would remark that current cosmological observations (dark energy and dark matter issues) are definitely taking up this role.}, we definitely need it, not only on philosophical grounds (reductionism as a driving force in physics), but also because we know that in physically relevant regimes (e.g.~singularities in cosmology and in black holes...) our classical theory of gravitation fails to be predictive. However, when searching for QG, we have to tackle not only deep theoretical problems (e.g.~the renormalizability of gravitational theories, the possible loss of unitarity in gravitational phenomena \cite{Hawking:1976ra}, the meaning of time in QG \cite{Isham:1995wr,Butterfield:1998dd})  but also the lack of observational and experimental guidance. The typical scale at which QG effects should become relevant is expected to be the one at which the gravitational action (the Einstein-Hilbert action for General Relativity) becomes of the order of the quantum of action $\hbar$. This happens at the so called Planck scale $\Mpl \equiv \sqrt{\hbar c/G_{N}}\simeq 1.22\times 10^{19}~\GeV/c^{2}$ which corresponds to energies well above the capabilities of any Earth based experiment as well as any observationally accessible regime.

However, the situation may be better than it appears at first sight. In fact, models of gravitation beyond General Relativity and models of QG have shown that there can be several low energy ``relic signatures'' of these models, which would lead to deviation from the standard theory predictions in specific regimes. 
%

Here we focus upon the phenomenology of violations of fundamental symmetries, given that a convenient way to perform high-precision tests is to look for experimental deviations from symmetries that are believed to hold {\em exactly} in nature and that could be broken by QG. 

An example of such a fundamental symmetry is CPT invariance, which requires that physics be unchanged under the combination of charge conjugation (C), parity inversion (P) and time reversal (T). C connects particles and antiparticles, P represents a spatial reflection of physical quantities with respect to the coordinate origin and T reverses a physics reaction in time.

In Quantum Field Theory, Lorentz symmetry is intimately related to CPT symmetry. Indeed, one of the hypotheses of the well known ``CPT theorem'' is Lorentz invariance. If CPT is broken, then at least one of the hypotheses of the CPT theorem should also break down. It has been proven \cite{Greenberg:2002uu} that Lorentz symmetry is the failing assumption in the so called ``anti-CPT theorem'', which states that in any unitary, local, relativistic point-particle field theory CPT breaking implies Lorentz violation. Note however that the converse of this statement is not true: it is possible to violate Lorentz invariance while keeping CPT exact\footnote{However, this theorem does not hold for theories that do not admit a local field theory formulation and that can therefore have unexpected properties \cite{Chaichian:2011fc}.}.

Thus, it is interesting to study both the theory and the phenomenology of Lorentz invariance violation (LV), which may yield a glimpse of QG. 
%
%
In recent years, attempts to place constraints on high-energy deviations from LI have mainly focused on modified dispersion relations for elementary particles.  Indeed, specific hints of LV arose from various approaches to Quantum Gravity. Among the many examples are string theory tensor VEVs \cite{KS89}, space-time foam~\cite{AmelinoCamelia:1997gz}, semiclassical spin-network calculations in Loop QG~\cite{Gambini:1998it}, non-commutative geometry~\cite{Carroll:2001ws, Lukierski:1993wx, AmelinoCamelia:1999pm}, some brane-world backgrounds~\cite{Burgess:2002tb} and condensed matter analogues of ``emergent gravity''~\cite{Barcelo:2005fc}. 

Lorentz symmetry breaking is not a necessary feature of QG, but it is clear that any possible LV effect connected with the Planck scale could provide an observational window into QG. However, to directly observe phenomena connected with $\Mpl$ would require the center of mass energy of, e.g., a scattering process to be comparable to $\Mpl$. This is 15 orders of magnitude larger than what the LHC can probe with its design center of mass energy of $14~\TeV$. On the other hand, if we are testing LI specifically, then also non-LI quantities can be important. The energy of the particle in some frame, or a cosmological propagation distance are widely discussed examples. These quantities can be so large as to effectively offset the $\Mpl$ suppression to a physical observable, so that very small corrections are magnified. For this reason, they are called ``windows on QG''. 

In order to correctly identify such ``windows on QG'' it is important to place them into a dynamical framework. A standard method is to study, within the context of Effective Field Theory (EFT), a Lagrangian containing the standard model fields and all LV operators of interest that can be constructed by coupling the standard model fields to new LV tensor fields that have non-zero vacuum expectation values \cite{Colladay:1998fq,Myers:2003fd,GrootNibbelink:2004za,Bolokhov:2007yc,Mattingly:2008pw,Kostelecky:2009zp}.~\footnote{There are other approaches to either violate or modify Lorentz invariance, that do not necessarily yield a low energy EFT (see \cite{AmelinoCamelia:2008qg} and refs therein). However, these models do not easily lead to particle physics constraints as the dynamics of particles is less well understood. Therefore we do not consider them here. In particular, we remark here that ideas of deformation, rather than breaking, of the Lorentz symmetry (see, e.g., \cite{AmelinoCamelia:2000mn}) do not have an ordinary-EFT formulation, hence they cannot be tested with the procedures discussed here.} A generic result of this procedure is the presence of modified dispersion relations for particles, of the form
\begin{equation}
E^{2}-p^{2} = m^{2} + f(\vec{p},\Mpl; \mu)\;,
\label{eq:genericmdr}
\end{equation}
where $m$ is the particle mass, $E$ its energy, the function $f$ represents the QG contribution and can depend generically on the momentum $\vec{p}$, on $\Mpl$ and on some intermediate mass scale $\mu$. For simplicity we assume that only boost invariance is broken, while rotations are preserved (see \cite{Mattingly:2005re} for further comments on rotation breaking), so that $f$ depends on $p=|\vec{p}|$, rather than on $\vec{p}$. Moreover, at $p \ll \Mpl$ we can expand $f$ so that Eq.~(\ref{eq:genericmdr}) reads
\begin{equation}
E^{2}-p^{2} = m^{2} + \sum_{n=1}^{N}\eta^{(n)} \frac{p^{n}}{\Mpl^{n-2}}\;,
\label{eq:mdrgeneral}
\end{equation}
%

All renormalizable LV operators (corresponding to $n=1,2$) that can be added to the standard model
are known as the (minimal) Standard Model Extension
(mSME)~\cite{Colladay:1998fq}. These operators all have dimension
three or four and can be further classified by their behavior under
CPT. The CPT odd dimension five kinetic terms for QED were written
down in~\cite{Myers:2003fd} while the full set of dimension five
operators were analyzed in~\cite{Bolokhov:2007yc}. The dimension
five and six CPT even kinetic terms for QED for particles coupled to
a non-zero background vector, which we are primarily interested in
here, were analyzed in~\cite{Mattingly:2008pw}. It is notable that
SUSY forbids renormalizable operators for matter coupled to non-zero
vectors~\cite{GrootNibbelink:2004za} but permits nonrenormalizable
operators.

Many of the operators in these various EFT parameterizations of LV
have been very tightly constrained via direct observations (see~\cite{Mattingly:2005re, Jacobson:2005bg, Liberati:2009pf} for extensive reviews).  Moreover, higher dimension LV operators can be tightly constrained by EFT arguments~\cite{Collins:2004bp}
showing that they will generically induce via radiative corrections large dimension 3 and 4 
operators in coupled particles if we assume no other relevant physics enters
between the TeV and $\Mpl$ energies. 

This is a very powerful argument which applies basically to any Lorentz violating theory (see e.g.~\cite{Iengo:2009ix}) and should not be arbitrarily discounted. However, as the SUSY example above shows, this assumption can be a little dangerous
as new physics above a TeV can change the hierarchy of terms. In particular SUSY would prohibit dimension 3 and 4 
operators and once broken would add an extra $\mathcal{O}(\Lambda_{SUSY}/\Mpl)$-to-some-suitable-power suppression. We still do not know if SUSY can really do this job and it is not clear if this is the correct solution of this naturalness problem. However, given the present uncertainty on this, it would be nice, when possible, to constrain the dimension five and six LV kinetic terms directly via observation. 

Mass dimension five CPT odd operators have been strongly constrained, both in QED and in the hadronic sector, using a wealth of observations spanning from the synchrotron spectrum of the Crab nebula (and its hard X-ray polarization) to the ultra-high-energy cosmic rays (UHECR), (see e.g.~\cite{Liberati:2009pf}). Here we shall consider explicitly terms coming from dimension five and six CPT even LV operators, so that the dispersion relations for protons, pions, and photons respectively, take the form~\cite{Maccione:2009ju}
\begin{equation}
E_p^2=p^2  + m_p^2 +  \eta_p \frac {p^4} {\Mpl^2},\quad E_\pi^2=p^2  + m_\pi^2+ \eta_\pi \frac {p^4} {\Mpl^2},\quad \omega^2=k^2  +  \beta \frac {k^4} {\Mpl^2}\label{eq:finaldisp}
\end{equation}
The successful operation of the PAO has brought UHECRs to the interest of a wide community of scientists and has allowed to test fundamental physics (in particular Lorentz invariance in the QED sector) with unprecedented precision \cite{PhysRevLett.100.021102,Maccione:2008iw,Galaverni:2008yj}. Given that the most important developments in the last years were achieved in the context of UHECR physics, we will review the role of UHECRs in the following.

\section{Ultra-high-energy Cosmic Rays and LV}

The UHECR constraints \cite{Kifune:1999ex,Aloisio:2000cm,AmelinoCamelia:2000zs,Stecker:2004xm,GonzalezMestres:2009di,Scully:2008jp,Maccione:2009ju,Stecker:2009hj} rely on the behavior of particle reaction thresholds with LV. What matters for threshold reactions in the presence of modified dispersion relations as in Eq.~(\ref{eq:mdrgeneral}) is not the size of the LV correction compared to the absolute energy of the particle, but rather the size of the LV correction to the mass of the particles in the reaction.  Hence the LV terms usually become important when their size becomes comparable to the mass of the heaviest particle. This criterion sets the presence of a critical energy $E_{cr}$ above which LV effects are relevant in a given threshold reaction. If the LV term scales with energy as $E^{n}$, then $E_{cr} \sim \left(m^{2}\Mpl^{n-2}\right)^{{1}/{n}}$ \cite{Jacobson:2002hd}. According to this reasoning, the larger the particle mass the higher is the energy at which threshold LV effects come into play. 


\subsection{Constraints from the UHECR spectrum}

The Cosmic Ray spectrum spans more than ten decades in energy (from
$<100~\MeV$ to $>10^{20}~\eV$) with a power-law shape of impressive
regularity ${dN}/{dE} \propto E^{-p}$.
The spectral slope $p$ has been measured as $p \simeq 2.7$ for $1~\GeV \lesssim E \lesssim 10^{15.5}~\eV$, followed by a softening (the ``knee'') to $p \simeq 3.0$ for $10^{15.5}~\eV \lesssim E \lesssim 10^{17.5}~\eV$, a further steepening to $p\simeq 3.2$ (the ``second knee'') up to $E \simeq 10^{18.5}~\eV$ and a subsequent hardening (the ``ankle'') to again $p \simeq 2.7$ at $E \gtrsim 10^{18.5}~\eV$ \cite{Gaisser:2006sf,Bergman:2007kn}.

One of the most fascinating problems regarding CRs is at what energy the end-point to the
CR spectrum occurs. A suppression to the spectrum is expected theoretically due to the interactions of UHECR protons with the Cosmic Microwave Background (CMB), leading to the production of charged and neutral pions, eventually dumping the energy of the UHECR protons into neutrinos and $\gamma$-rays. This reaction has a LI threshold energy $E_{\rm th} \simeq 5\times 10^{19}~(\omega_{b}/1.3~\meV)^{-1}~\eV$ ($\omega_{b}$ is the target photon energy). Therefore, at the present epoch, significant photo-pion production in a LI theory occurs only if the energy of the interacting proton is above a few $10^{19}~\eV$, with the CR mean-free-path rapidly decreasing above this energy. Hence, it has long been thought to be responsible for a cut-off in the UHECR spectrum, the Greisen-Zatsepin-Kuzmin (GZK) cut-off \cite{JETPLetters.4.3.78,Greisen:1966jv}. Moreover, trans-GZK particles arriving at Earth must be accelerated within the so called GZK sphere, whose radius is
expected to be of the order of 100~Mpc at $\sim 10^{20}~\eV$ and to
shrink down at larger energies. 
Experimentally, the presence of a suppression of the UHECR spectrum
has been confirmed only recently with the observations by the HiRes
detector \cite{Abbasi:2007sv} and the PAO
\cite{Roth:2007in}. Although the cut-off could be also due to the
finite acceleration power of the UHECR sources, the fact that it
occurs at roughly the expected energy favors a GZK explanation. The
correlation results shown in \cite{Cronin:2007zz} further strengthen this
hypothesis. 

It is this scenario where possible LV effects come into play. 
LV has two effects on UHECR propagation: it modifies standard reactions and allows new, normally forbidden reactions. In particular, in \cite{Maccione:2009ju} it was considered 
\begin{itemize}
\item $p+\gamma \rightarrow p+\pi^{0} ~(n+\pi^{+})$, which is modified by LV.
\item $p\rightarrow p+\gamma$ and  $p\rightarrow p+\pi$, which correspond respectively to photon and pion emission in vacuum and would be forbidden if LI were exact.
\end{itemize}
As a consequence of LV, the mean free path for the GZK reaction is modified. The propagated UHECR spectrum can therefore display features, like bumps at specific energies, suppression at low energy, recovery at energies above the cutoff, such that the observed spectrum cannot be reproduced. Moreover, the emission of Cherenkov $\gamma$-rays and pions in vacuum would lead to sharp suppression of the spectrum above the relevant threshold energy. After a detailed statistical analysis of the agreement between the observed UHECR spectrum and the theoretically predicted one in the presence of LV and assuming pure proton composition, the final constraints implied by UHECR physics are (at
99\% CL) \cite{Maccione:2009ju}
\begin{eqnarray}
\nonumber
-10^{-3} \lesssim &\eta_{p}& \lesssim 10^{-6}\\
-10^{-3} \lesssim &\eta_{\pi}&  \lesssim 10^{-1} \quad (\eta_{p} > 0) \quad\mbox{or}\quad \lesssim 10^{-6} \quad (\eta_{p} < 0)\,.
\label{eq:finalconstraint}
\end{eqnarray}
%


\subsubsection{Role of UHE nuclei}
UHECR constraints have relied so far on the hypothesis, not in contrast with any previous experimental evidence, that protons constituted the majority of UHECRs above $10^{19}~\eV$. Recent PAO \cite{Abraham:2010yv} and Yakutsk \cite{Glushkov:2007gd} observations, however, showed strong hints of an increase of the average mass composition with rising energies 
up to $E \approx 10^{19.6}~\eV$, although still with large uncertainties. Hence, experimental data suggests that heavy nuclei can possibly account for a substantial fraction of UHECR on Earth.

One can assume that each individual nucleus has its own independent modified dispersion relation and make a further simplification by assuming that energy and momentum of the nucleus are the sum of energies and momenta of its constituents \cite{Jacobson:2002hd}. With this approximation (and also taking the masses of protons and neutrons to be equal), the dispersion relation for a nucleus of mass $A$ and charge $Z$ can be written as \cite{Saveliev:2011vw}
\begin{equation}
E_{A}^{2} = \left( A E_{1} \right)^{2} = \left( A p_{1} \right)^{2} + \left( A m_{1} \right)^{2} + \frac{\eta}{A^{2}}\frac{\left( A p_{1} \right)^{4}}{\Mpl^{2}} = p^{2}_{A,Z} + m^{2}_{A,Z} + \frac{\eta_{p}}{A^{2}}\frac{p^{4}_{A,Z}}{\Mpl^{2}}\,.
\label{eq:mdr}
\end{equation} 
So now we have only one free parameter, $\eta_{p}$, for the nucleon, while for nuclei there are effective parameters of the form $\eta_{A} = \eta_{p}/A^{2}$. This phenomenological model guarantees that the correct dispersion relations are recovered when dealing with macroscopic objects \cite{Jacobson:2002hd}, for which QG effects should be suppressed.

Assuming that current hints for a heavy composition at energies $E \sim 10^{19.6}~\eV$ \cite{Abraham:2010yv} may be confirmed in the future, and that some UHECR is observed up to $E \sim 10^{20}~\eV$ \cite{Abraham:2010mj}, one could place a first constraint on the absence of spontaneous decay for nuclei which could not spontaneously decay without LV \cite{Saveliev:2011vw}. It will place a limit on $\eta_{p}<0$, because in this case the energy of the emitted nucleon is lowered with respect to the LI case until it ``compensates'' the binding energy of the nucleons in the initial nucleus in the energy-momentum conservation.

An upper limit for $\eta_{p}>0$ can instead be obtained from vacuum Cherenkov emission \cite{Saveliev:2011vw}.  Assuming UHECR to be mainly iron at the highest energies the constraint is given by $\eta_{p} \lesssim 2\times10^{2}$ for nuclei observed at $10^{19.6}~\eV$ (and $\eta_{p} \lesssim 4$ for $10^{20}~\eV$), while for He it is $\eta_{p} \lesssim 4\times10^{-3}$ ($10^{-4}$). 

UHE nuclei suffer mainly from photodisintegration losses as they propagate in the intergalactic medium. Because photodisintegration is indeed a threshold process, it can be strongly affected by LV. According to \cite{Saveliev:2011vw}, and in the same way as for the proton case, the mean free paths of UHE nuclei are modified by LV in such a way that the final UHECR spectra after propagation can show distinctive LV features. However, a quantitative evaluation of the propagated spectra has not been performed yet.

\subsection{Constraints from UHE $\gamma$-rays}

Photopion interactions of UHECR protons with the CMB lead to the production of neutral pions which subsequently decay into UHE $\gamma$-ray pairs. The PAO and the Yakutsk and AGASA experiments imposed limits on the presence of photons in the UHECR spectrum. In particular, the photon fraction is less than 2.0\%, 5.1\%, 31\% and 36\% (95\% C.L)~at $E = 10$, 20, 40, 100 EeV  respectively \cite{Aglietta:2007yx,Rubtsov:2006tt}. From the theoretical side, and bearing in mind the uncertainties related to source and propagation effects, it is well established that in a LI framework UHE photons are attenuated by pair production onto the CMB and Radio background during their travel to Earth, leading to their fraction in the total UHECR flux being reduced to less than 1\% at $10^{19}\eV$ and less than 10\% at $10^{20}~\eV$ \cite{Sigl:2007ea, Gelmini:2007jy}. It was shown in a framework with modified dispersion relations for both photons and $e^{+}/e^{-}$  and standard energy/momentum conservation, that pair production could be effectively inhibited at high energy, due to the presence of an upper threshold \cite{Galaverni:2007tq},\footnote{An upper threshold is an energy above which it is not possible to simultaneously conserve energy and momentum in an interaction.  If Lorentz symmetry is exact then upper thresholds do not exist, while they might well exist if it is violated \cite{Mattingly:2002ba}.} and therefore the fraction of photons present in UHECRs on Earth would violate the present experimental upper limits. Hence, the {\em non} observation of a large fraction of UHE photons in UHECRs implies the constraint $|\xi| < \mathcal{O}(10^{-14})$ in the EFT framework \cite{Maccione:2008iw,Galaverni:2008yj}. 

\subsubsection{Constraints on space-time foam models}
The recent detection of time delays on arrival of high energy $\gamma$-rays \cite{Abdo:2009zz,Ackermann:2009zq} led to renewed interest of the astrophysics community in QG induced LV effects. The observed time delays can be explained, and are actually expected, in standard astrophysical scenarios hence they can be readily used to place constraints on LV models. However, time delays are naturally predicted also in generic LV QG models. It is now established that any LV model able to reproduce the observed delays and admitting an EFT formulation is in tension with other astrophysical observations (see e.g.~\cite{Liberati:2009pf}). Up to now, the only fully developed LV model able to explain the observed time delays has a string theory origin and does not admit an EFT formulation \cite{Ellis:1992eh,Ellis:2000sf,Ellis:2003if,Ellis:2003sd,Ellis:2008gg, Ellis:2009yx, Li:2009tt, Ellis:2009vq}. Therefore, if observed time delays were due to such QG effects, the propagation of GeV photons over cosmological distances could not be described within EFT. Given that EFT is accurately verified with terrestrial accelerators up to $\sim 100~\GeV$, this would be a very striking and revolutionary conclusion.

In the model \cite{Ellis:1992eh,Ellis:2000sf,Ellis:2003if,Ellis:2003sd,Ellis:2008gg, Ellis:2009yx, Li:2009tt, Ellis:2009vq} only purely neutral particles, such as photons or Majorana neutrinos, possess LV modified dispersion relations. For photons this has the form 
\begin{equation}
E_{\gamma}^{2} = p^{2} - \xi \frac{p^{3}}{M}\;,
\label{eq:mdrDb}
\end{equation}
with the free parameter $\xi>0$. Hence only subluminal photons are present in the theory, and photon propagation in vacuum is not birefringent. Due to stochastic losses in interactions with the D-brane foam, exact energy-momentum conservation during interactions does not hold. This last phenomenon is controlled by the free parameter $\xi_{I}$ \cite{Ellis:2000sf}.

According to Eq.~(\ref{eq:mdrDb}) photons with different energy travel at different speeds. Then, if a source at redshift $\bar{z}$ simultaneously emitted two photons at energy $E_1' \neq E_2'$, their time delay at Earth will be
\begin{equation}
\Delta t \simeq \xi \frac{\Delta E}{M}\frac{1}{H_{0}}\int_{0}^{\bar{z}}dz\frac{1+z}{\sqrt{\Omega_{\Lambda} + (1+z)^{3}\Omega_{\rm M}}}\;,
\label{eq:tof}
\end{equation}
where $\Delta E$ is the observed energy difference and the integral on redshift accounts also for redshift of the energy \cite{Jacob:2008bw,AmelinoCamelia:2009pg,Ellis:1999sd}. Time-of-flight constraints are then viable for this model, even though they lead at most to constraints on $\xi$, because $\xi_{I}$ is not effective in this context. 

Rather intriguingly, the FERMI Collaboration has recently reported the detection of delays on arrival of $\gamma$-ray photons emitted by distant GRBs, in particular GRB 080916C \cite{Abdo:2009zz} and GRB 090510 \cite{Ackermann:2009zq} (see however \cite{AmelinoCamelia:2009pg} for an updated review). A thorough analysis of these delays in the energy range 35 MeV -- 31 GeV allowed to place for the first time a conservative constraint of order $\xi \lesssim 0.8$ \cite{Ackermann:2009zq} on LV effects expressed as in Eq.~(\ref{eq:mdrDb}). 
This is the best constraint so far available on the theory. On the other hand, FERMI results can be interpreted in terms of LV assuming $\xi\simeq 0.4$ and a possible evolution of the D-particle density with redshift~\cite{Ellis:2009vq}.\footnote{Plausible astrophysical explanations of this phenomenon exist. No claim of a discovery of LV can be made on the basis of the data reported in \cite{Abdo:2009zz, Ackermann:2009zq}, where only LV constraints are discussed.}

In order to constrain the D-brane model, the process of pair production, $\gamma\gamma \rightarrow e^{+}e^{-}$, can be exploited \cite{Maccione:2010sv}. Indeed, according to \cite{Maccione:2010sv} also in the D-brane model pair-production exhibits upper thresholds which, for values of the free parameters $\xi,\xi_{I} \gtrsim \mathcal{O}(10^{-12})$, are located at $E>10^{19}~\eV$. This would lead to UHECRs being constituted by a large fraction of photons, in contrast with experimental data \cite{Maccione:2010sv}.
Therefore, a limit $\xi,\xi_{I} \lesssim 10^{-12}$ is placed \cite{Maccione:2010sv}. This also means that D-particle explanations of time delays in the GeV range are in conflict with data on the photon fraction in UHECRs (although some possible implementations of the model \cite{Ellis:2008gg} were recently proposed \cite{Ellis:2010he} which would naturally evade the above constraints).

\subsection{Foreseen constraints from UHE neutrinos}

Neutrinos, with their tiny mass of order $m_{\nu} \simeq 0.01~\eV$ \cite{pdg}, are in principle the most suited particles to provide strong constraints on LV, at least for reactions involving {\em only} neutrinos.  
%
Despite the threshold being low for LV effects to kick in, neutrinos with ultra-high energy are necessary to achieve a signal, as they interact so weakly that the phase space for a LV reaction must be huge to generate an appreciable rate. This requirement implies that very large energies are needed. 


If one neglects exotic sources of UHE neutrinos, the ``cosmogenic'' neutrino flux is created \cite{Beresinsky:1969qj,Stecker:1973sy,Engel:2001hd,Semikoz:2003wv} via the decay of charged pions produced by the interaction of primary nucleons with CMB photons above the GZK threshold.  Violation of LI however introduces new phenomena in the propagation of UHE neutrinos. A detailed list can be found in \cite{Mattingly:2009jf}, however we shall focuss here on the so called $\nu$-splitting $\nu\rightarrow  \nu \nu\nubar$ as it exclusively involves the neutrino sector and has a high enough rate to be seen at UH energies.

The effects of neutrino splitting on the UHE neutrino spectrum are
twofold and can be understood qualitatively as follows.
\begin{description}
\item[Flux suppression at UH energies] The splitting is effectively an energy loss process for UHE neutrinos. If the rate is sufficiently high, the energy loss length can be below 1 Mpc. Let us call $\bar{E}(\eta_\nu)$ the energy at which this happens. Then, being GZK neutrinos produced mainly at distances larger than 1 Mpc, we do not expect any neutrino to be detected at Earth with $E>\bar{E}$.
The mere observation of neutrinos up to a certain energy $E_{\rm obs}$ would
imply a constraint \cite{Mattingly:2009jf}
\begin{equation}
\eta_{\nu}^{(4)} \lesssim
\left(\frac{E_{\rm obs}}{6\times10^{18}~\eV}\right)^{-13/4}\;.
\label{eq:constraint_naive}
\end{equation}

\item[Flux enhancement at sub-UH energies] Neutrinos lose energy by producing lower energy neutrinos. Eventually these neutrinos will become stable, either because their energy is below threshold, or because their lifetime is larger than their propagation time. Accordingly, an enhancement of the neutrino flux at energies below ${\rm few} \times 10^{18}~\eV$ is expected \cite{Mattingly:2009jf}. 
\end{description}

Next generation neutrino detectors such as ANITA \cite{Gorham:2008yk} and SuperEUSO \cite{Petrolini:2009cg,Santangelo:2009et} are sensitive to neutrinos of energies $>10^{19}~\eV$. Further experiments, like the planned ARIANNA \cite{Barwick:2006tg,Barwick:2009zz} and IceRay \cite{Allison:2009rz}, will cover the range $10^{17}\div 10^{20}~\eV$. The scenarios described above can then be tested in the near future and constraints $\eta_{\nu} < 10^{-4}$ will be potentially cast according to Eq.~(\ref{eq:constraint_naive}).

\section{Summary}

QG phenomenology of Lorentz and CPT violations is a success story in physics. We have progressed in few years from almost no tests to tight, robust constraints on EFT models and some spacetime foam models. 
In summary for EFT with LV the situation is:
\begin{description}
\item[QED] up to $\mathcal{O}(10^{-22})$ on $n=2$, $\mathcal{O}(10^{-11})$ on $n=3$, $\mathcal{O}(10^{-7})$ on $n=4$
\item[Hadrons]  up to $\mathcal{O}(10^{-50})$ on $n=1$, $\mathcal{O}(10^{-27})$ on $n=2$, $\mathcal{O}(10^{-14})$ on $n=3$, $\mathcal{O}(10^{-6})$ on $n=4$
\item[Neutrinos] up to $\mathcal{O}(10^{-27})$ on $n=2$, $\mathcal{O}(10^{-14})$ on $n=3$, expected $\mathcal{O}(10^{-4})$ on $n=4$
\end{description}
Chances are high that improving observations in HE astrophysics will strengthen these constraints in a near future. 
Let us note however, that there is a noticeable missing voice in the above list, this is the gravitational sector. In particular, it would be important to cast in the future constraints which are purely gravitational, given that this framework seems to have both theoretical \cite{Horava:2009uw} as well as phenomenological reasons for being pursued \cite{Pospelov:2010mp}\footnote{Basically if LV operators are present only in the gravitational sector the induced LV operators in Standard Models particles are further suppressed by the smallness of the Newton constant provided the Lorentz breaking scale is much lower energy scale than the Planck one. This could be then a way out of the previously discussed naturalness problem.}. We leave this for future studies.

\section{Conclusions and Perspectives}

Lorentz invariance of physical laws relies on only few assumptions: the principle of relativity, stating the equivalence of physical laws for non-accelerated observers, isotropy (no preferred direction) and homogeneity (no preferred location) of space-time, and a notion of precausality, requiring that the time ordering of co-local events in one reference frame be preserved \cite{brown2005physical,Liberati:2001sd,Sonego:2008iu}.
In this sense a breakdown of Lorentz invariance does not necessarily imply a breakdown of the relativity principle. For this reason, it is worth exploring an alternative possibility that keeps the relativity principle but that relaxes one or more of the above postulates. Such a possibility can lead to the so-called very special relativity framework \cite{Cohen:2006ky}, which was discovered to correspond to the break down of isotropy and to be described by a Finslerian-type geometry~\cite{Bogoslovsky:2005cs,Bogoslovsky:2005gs,Gibbons:2007iu}.  Notice that in this example the generators of the new relativity group number fewer than the usual ten associated with Poincar\'e invariance. Specifically, there is an explicit breaking of the $O(3)$ group associated with rotational invariance. Finsler-type geometries have also been considered as a possible geometric framework for modified dispersion relations like Eq.~(\ref{eq:mdrgeneral}) in \cite{Girelli:2006fw} albeit the possibility to use them as the geometric counterpart of Minkowski spacetime for alternative special relativity groups seems hampered by structural problems (see e.g.~\cite{Skakala:2010hw} and references therein).

One may wonder whether there exist alternative relativity groups with the same number of generators as special relativity. Currently, we know of no such generalization in {(commutative)} coordinate space. However, it has been suggested that, {in non-commutative spacetime}, such a generalization is possible, and it was termed  ``doubly" or ``deformed" (to stress the fact that it still has 10 generators) special relativity, DSR~\cite{AmelinoCamelia:2011bm}. Unfortunately, the various DSR candidates  face in general major  problems regarding  their physical interpretation and a working model is not yet available (see however \cite{AmelinoCamelia:2011bm} for recent attempts in new directions). 

Finally, it is a {logical, and rather simple, possibility that  a Lorentz symmetry breakdown} could be signaling an interpolation from a relativity group to another one, for example two special relativity groups characterized by different limit speeds (see \cite{Fagnocchi:2010sn} for an example in the so called analogue gravity context \cite{Barcelo:2005fc}) or between a Lorentzian and an Euclidean Poincar\'e group (see \cite{Girelli:2008qp} for an explicit, analogue gravity inspired, example). Even more intriguingly it might be that a Lorentz invariant world could emerge from a non-relativistic system living in lower dimensions (e.g.~the effective dimension of quantum gravity models seems to generically reduce to two at very short scales \cite{Carlip:2009kf})

In conclusion, we should take the experience in constraining EFT with LV as a lesson that we can and we must challenge quantum/emergent gravity scenarios with the observational test. However, we cannot say yet ``mission accomplished" for what regards testing possible deviations from local Lorentz invariance at small scales as we have started testing the most obvious, generic scenarios. New tests will probably require picking up more specific models for what lies beyond the Planck scale and will rely more heavily on these assumptions. This task will also probably require a better use and knowledge of the current (mainly astrophysical) data. Still we feel that any possibility to confront our ideas with reality should be pursued without hesitations, and that the path walked in these years should be followed. There cannot be any credible quantum gravity research without a rigorous quantum gravity phenomenology challenge.

\section*{References}
\bibliographystyle{iopart-num}
\bibliography{references}

\end{document}